\documentclass[twocolumn,english]{revtex4}
\usepackage[T1]{fontenc}
\usepackage[latin1]{inputenc}
\usepackage{graphicx}

\makeatletter

\usepackage{babel}
\makeatother
\begin{document}

\title{Incipient magnetism in the cubic perovskites $MgCNi_{3}$ and $YBRh_{3}$:
A comparison}

\author{P. Jiji Thomas Joseph and Prabhakar P. Singh}

\address{Department of Physics, Indian Institute of Technology Bombay, Powai,
Mumbai-400076, India }

\begin{abstract}
Using density-functional-based methods, we have studied the effects
of incipient magnetism in the cubic perovskites $MgCNi_{3}$ and $YBRh_{3}$.
Our results show that (i) at the equilibrium volume, both $MgCNi_{3}$
and $YBRh_{3}$ alloys remain paramagnetic and (ii) at expanded volumes,
only $YBRh_{3}$ shows the possibility of a ferromagnetic phase with
a local magnetic moment larger than $0.25\,\mu_{B}$ per $Rh$ atom. 
\end{abstract}
\maketitle
The coexistence of magnetism and superconductivity, predicted to be
incompatible \cite{PR-106-162} due to pair-breaking effects of magnetic
scattering of electrons \cite{SovPhys-12-1243}, has been observed
in a wide variety of materials such as the cuprates \cite{ZPhys-64-189,PRL-58-908},
the ruthenates \cite{Nature-372-572,PRL-83-3713}, the heavy fermion
compounds \cite{Nature-406-587,PRL-43-1892}, the organic superconductors
\cite{SSC-33-1119,JPhysLett-41-L95}, the borocarbides \cite{PRL-72-274,Nature-367-254}
and the transition-metal intermetallics $ZrZn_{2}$ \cite{Nature-412-58}. 

One of the recently discovered superconducting material, the cubic
perovskite $MgCNi_{3}$ \cite{Nature-411-54} with a superconducting
transition temperature $T_{C}=8\, K$, shows subtle signatures of
incipient magnetism in the form of spin-fluctuations. The density
of states of $MgCNi_{3}$ has a strong van-Hove singularity, primarily
composed of $Ni$ $3d$ states, just below the Fermi energy $E_{F}$
\cite{PRB-66-172507,PRL-88-027001}, which suggests that the material
may be close to a magnetic instability. Using the rigid band approach,
one expects $\sim0.5$ holes to drive $MgCNi_{3}$ to ferromagnetism
\cite{PRB-64-100508,PRL-88-027001}. However, experiments find that
hole-doped $MgCNi_{3}$, with the hole-doping achieved via $Fe$ or
$Co$ substitutions in the $Ni$ sub-lattice \cite{PRB-66-064510,SSC-132-379,SSC-119-491,BJP-32-755,PRB-68-064503},
and vacancies \cite{SSC-121-73,PhyC-371-1} or $B$ substitutions
in the $C$ sub-lattice \cite{condmat-0412551}, remains non-magnetic. 

A variety of transition-metal-rich cubic perovskite materials have
been synthesized \cite{J.SSChem-1244} and tested for superconductivity.
Confirmed report of superconductivity appears only for $YB_{x}Rh_{3}$,
for which resistivity characterization finds $T_{C}$ of $\sim1$K
\cite{JLCM-125-233}. Magnetic measurements show that $YBRh_{3}$
remains a Pauli paramagnet over a wide temperature range, $4.2\leq T\leq300\, K$
\cite{JLCM-125-233}. The density of states of $YBRh_{3}$ reveals
that the density of states at $E_{F}$, $N(E_{F})$, is dominated
by the $Rh$ $4d$ states, with little or no contribution of the $B$
$2p$ states \cite{PRB-52-12921,JAC-349-206}. For expanded volumes,
the $E_{F}$ is expected to lower in energy. Such a movement in $E_{F}$
is expected to enhance $N(E_{F})$, thus raising the possibility of
satisfying the Stoner criteria. At the equilibrium volume, model calculations
for $MgCNi_{3}$ show that the Stoner criteria is far from being satisfied
\cite{PRL-88-027001,PRB-64-140507}, but is found large enough to
induce spin-fluctuations. 

Given that $MgCNi_{3}$ and $YBRh_{3}$ are perovskite superconductors,
and that incipient magnetism resides in $MgCNi_{3}$ in the form of
spin-fluctuations \cite{PRB-64-140507,PRL-87-257601}, it is interesting
to compare the propensity of magnetism in the two alloys as a function
of volume. The present study of ordered $MgCNi_{3}$ and $YBRh_{3}$
alloys is a step in that direction.

The unpolarized, spin polarized, and fixed-spin moment calculations
for ordered $MgCNi_{3}$ and $YBRh_{3}$ alloys are carried out, self-consistently,
using the Korringa-Kohn-Rostoker method in the atomic sphere approximation
(KKR-ASA). The calculations are scalar-relativistic with the partial
waves expanded up to $l_{max}=3$ inside the atomic spheres. The exchange-correlation
effects are taken into account using the Perdew and Wang parametrization
\cite{PRB-45-13244}. The core states have been recalculated after
each iteration. The overlap volume resulting from the blow up of the
muffin-tin spheres was less than $15$\%. The integration of the Green's
function over the energy to evaluate moments of the density of electronic
states was carried out along a semi circular contour comprising of
$20$ points in the complex plane. For the Brillouin zone integrations,
the special $\mathbf{k}-$point technique was employed with $1771$
$\mathbf{k}-$points spread in the irreducible wedge of the cubic
Brillouin zone. The convergence in the charge density was achieved
so that the root-mean square of moments of the occupied partial density
of states becomes smaller than $10^{-6}$.

The calculated equilibrium lattice constants for $MgCNi_{3}$ and
$YBRh_{3}$ are found to be $7.138\, a.u.$ and $7.919\, a.u.$, respectively.
For $MgCNi_{3}$ the calculated lattice constant is consistent with
the previous first-principles reports \cite{PRB-64-140507}, but the
value is underestimated by $1$\% when compared with the experiments
\cite{Nature-411-54}. Surprisingly, for $YBRh_{3}$ the calculated
lattice constant appears to be overestimated in comparison with the
experiments \cite{JLCM-125-233,JAC-354-12,JAC-291-52}, which is not
so common. A previous theoretical estimate of the lattice constant
of $YBRh_{3}$ also finds an overestimated value \cite{PRB-52-12921}.
Such a discrepancy also exists for $ZnCNi_{3}$ alloys \cite{Supercond-17-274,PRB-70-060507}.
For the latter, it is argued that the material $ZnCNi_{3}$ subjected
to experiments may be off-stoichiometric in $C$ sub-lattice \cite{PRB-70-060507}.
Thus, it can be inferred that the material $YBRh_{3}$, which is reported
to be a superconductor \cite{JLCM-125-233}, may be deficient in $B$
content, and if so, following the case of $MgCNi_{3}$ one may expect
higher $T_{C}$ for the stoichiometrically ordered materials.

\begin{figure}
\includegraphics[%
  clip,
  scale=0.35]{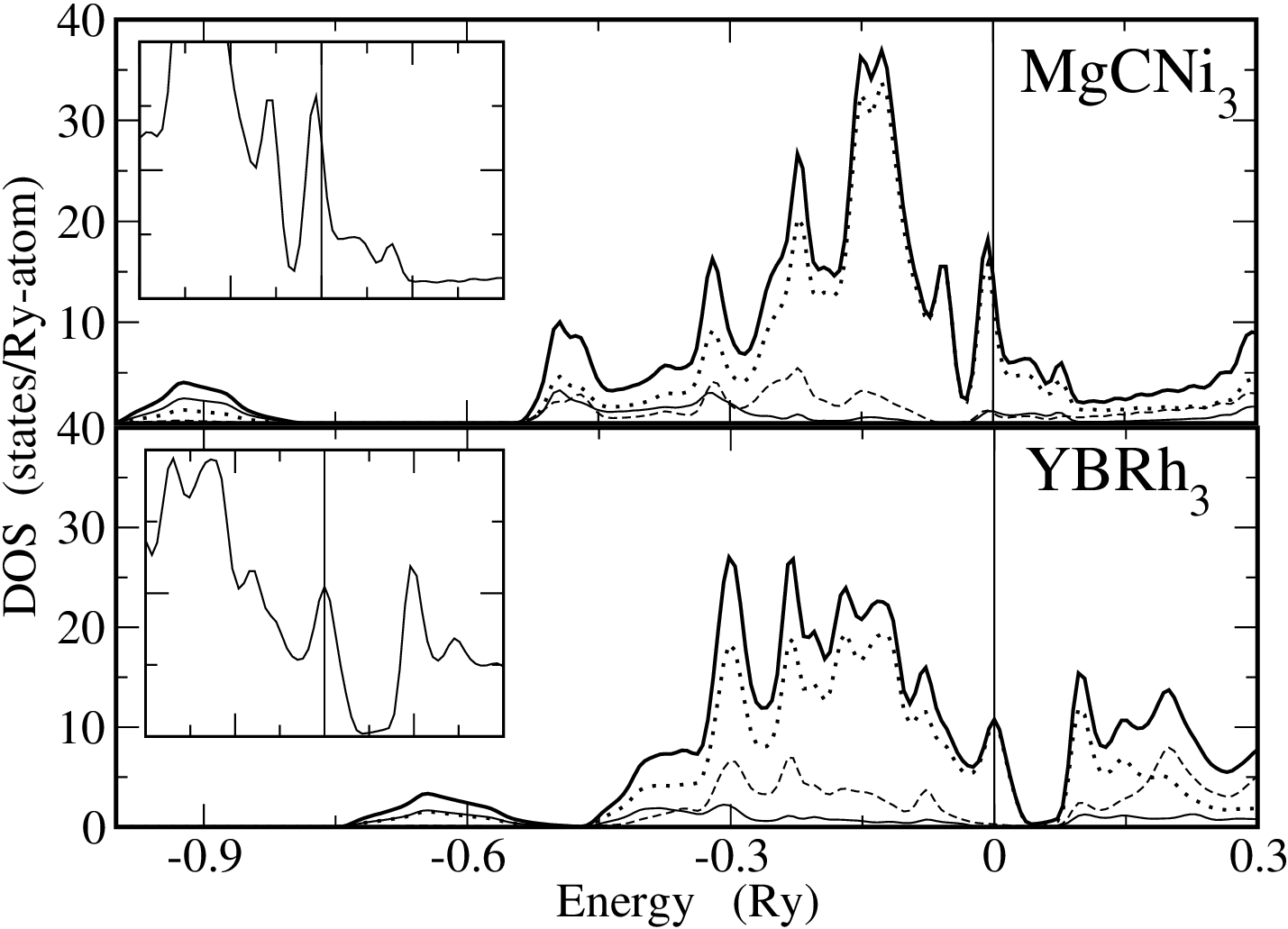}

\caption{\label{total-partial-dos}The total and sub-lattice resolved partial
densities of states of $MgCNi_{3}$ (upper panel) and $YBRh_{3}$
(lower panel) , calculated at their equilibrium lattice constants.
The thick line represents the total density of states while dotted,
dashed and thin lines represent partial densities of states for $Ni$
($Rh$), $Mg$ ($Y$), and $C$ ($B$). The inset is a blow up of
the states near the Fermi energy which is shown as a vertical line
through the energy zero.}
\end{figure}

The densities of states of $MgCNi_{3}$ and $YBRh_{3}$, calculated
at the respective equilibrium lattice constants, are shown in Fig.\ref{total-partial-dos}.
For both the materials, the density of states at $E_{F}$ is dominated
by the transition-metal states, $Ni$ in $MgCNi_{3}$ and $Rh$ in
$YBRh_{3}$. For $MgCNi_{3}$, the van-Hove singularity is well reproduced,
and it is about $0.06\, mRy$ below $E_{F}$, in agreement with the
earlier band-structure reports \cite{PRL-88-027001}. For $YBRh_{3}$,
the van-Hove singularity falls at $E_{F}$. In the itinerant model
of magnetism, a high $N(E_{F})$ is essential for magnetism, and thus
it appears that both $MgCNi_{3}$ and $YBRh_{3}$ alloys are close
to ferromagnetic instability. In the case of $YBRh_{3}$ the total
energies of both spin polarized and unpolarized calculations at the
equilibrium lattice constant remain degenerate, making $YBRh_{3}$
into a paramagnetic material. The calculated paramagnetic $N(E_{F})$
for $MgCNi_{3}$ and $YBRh_{3}$ are $14.562$ and $11.212\, state/Ry-atom$
respectively, which are consistent with the previous reports \cite{PRB-64-140507,PRL-88-027001,PRB-52-12921,PRB-64-100508,PRB-64-180510}. 

The characteristic difference in the densities of states of $MgCNi_{3}$
and $YBRh_{3}$ is the absence of non-metal $2p$ states at $E_{F}$
in $YBRh_{3}$. This compares well with its iso-electronic counterpart
$MgCCo_{3}$. However, $MgCCo_{3}$ is found to be a weak ferromagnet
with $Co$ having a local magnetic moment of $0.33\,\mu_{B}$ \cite{PRB-65-064525,JAP-91-8504}.
Next, we describe the effects of volume expansion on the incipient
magnetism in $MgCNi_{3}$ and $YBRh_{3}$. 

\begin{figure}
\includegraphics[%
  clip,
  scale=0.35]{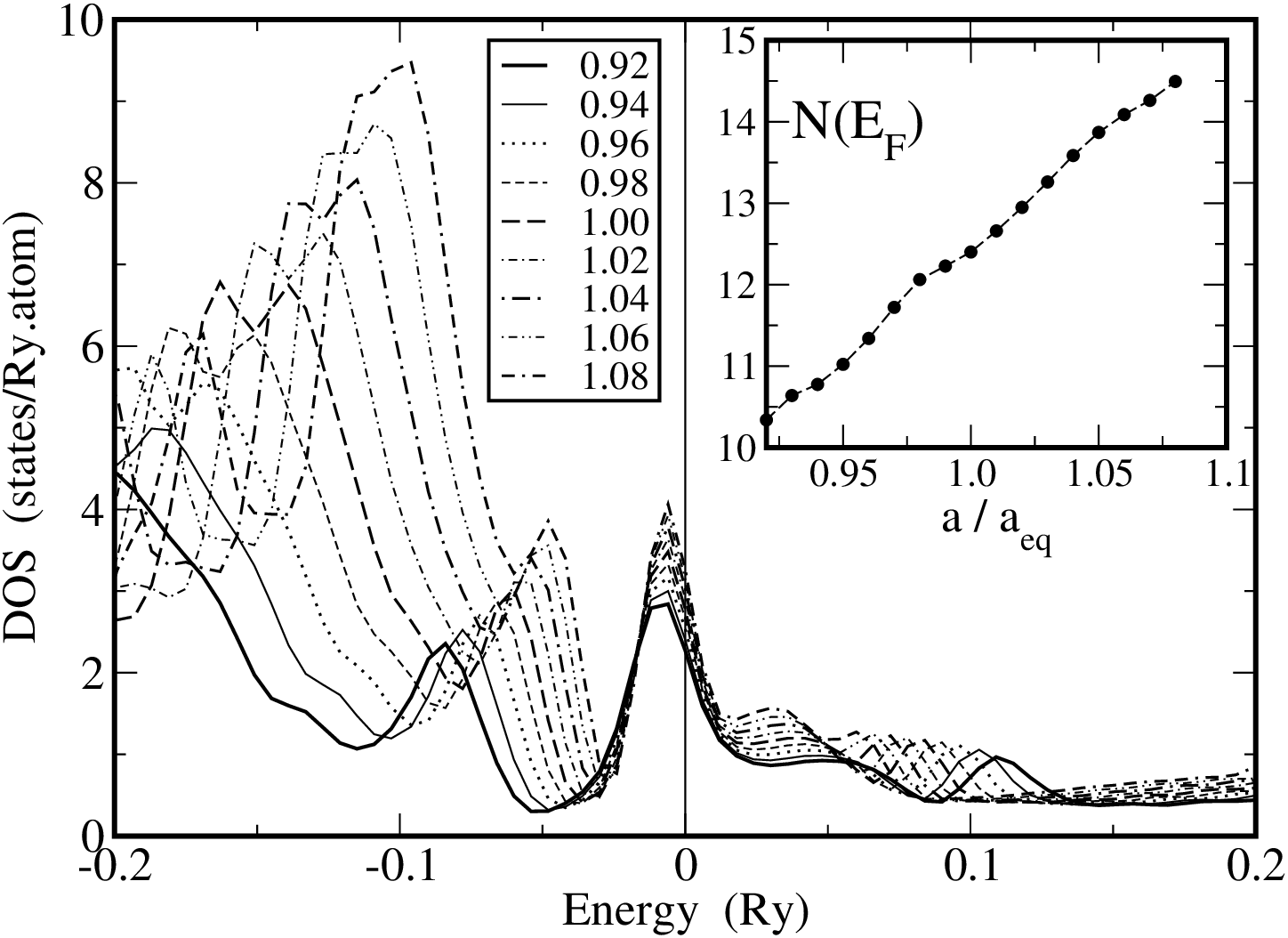}

\caption{\label{mgcni3-lat-dos}The total density of states of $MgCNi_{3}$
for various lattice constants (as indicated in the left inset), calculated
as described in the text. The density of states is shown over an energy
interval $-0.2\leq E\leq0.2$ with respect to the Fermi energy. The
vertical line through the energy zero represents the Fermi energy.
The total density of states at $E_{F}$ as a function of lattice constant
is shown in the right inset. }
\end{figure}

For $MgCNi_{3}$, up to a $15$\% increase in lattice constant from
the equilibrium showed no local magnetic moment at the $Ni$ site.
The results of our calculations are summarized in Fig.\ref{mgcni3-lat-dos},
where we show the density of states as a function of lattice constant
in $MgCNi_{3}$ over an energy interval of $-0.2\leq E\leq0.2$. Though
$N(E_{F})$ increases steadily with increasing lattice constant, the
peak, which is characteristic of the $Ni$ $3d$ bands, does not change
its position on the energy scale, consistent with the observations
made by Rosner $et.$ $al$. \cite{PRL-88-027001}. However, the increase
in $N(E_{F})$ is not sufficient to induce a magnetic phase transition
in $MgCNi_{3}$ even at expanded volumes. 

\begin{figure}
\includegraphics[%
  clip,
  scale=0.35]{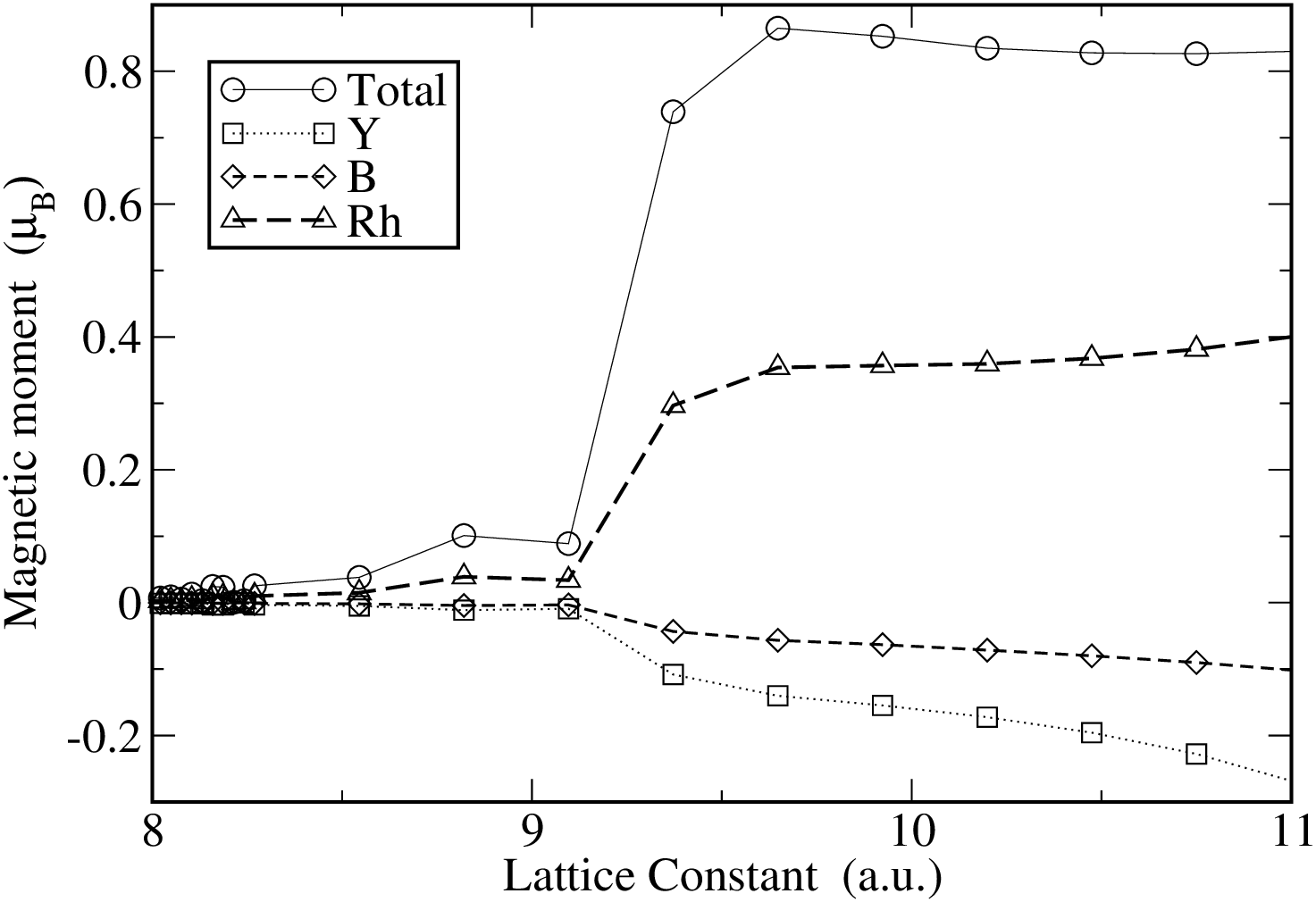}

\caption{\label{scf-loc-mom}The total magnetic moment as well as its contributions
from the $Y$, $B$ and $Rh$ sub-lattices in $YBRh_{3}$ as a function
of lattice constant, calculated as described in the text.}
\end{figure}

For $YBRh_{3}$, our calculations reveal a sharp ferromagnetic transition
with respect to expanded lattice constants. In Fig.\ref{scf-loc-mom},
we show the total magnetization as a function of lattice constant
for $YBRh_{3}$. The individual contributions to the magnetization
from $Y$, $B$ and $Rh$ sub-lattices in $YBRh_{3}$ are also shown
in Fig.\ref{scf-loc-mom}. We find a sudden appearance of local magnetic
moment larger than $0.25\,\mu_{B}$ per $Rh$ atom for lattice constants
larger than $15$\% of the equilibrium value in $YBRh_{3}$. Small
magnetic moments, which are anti-ferromagnetically coupled to the
magnetic moment at the $Rh$ site, also develop at the $Y$ and $B$
sites due to the hybridization with the $4d$ states of $Rh$. 

\begin{figure}
~

~

~

\includegraphics[%
  scale=0.35]{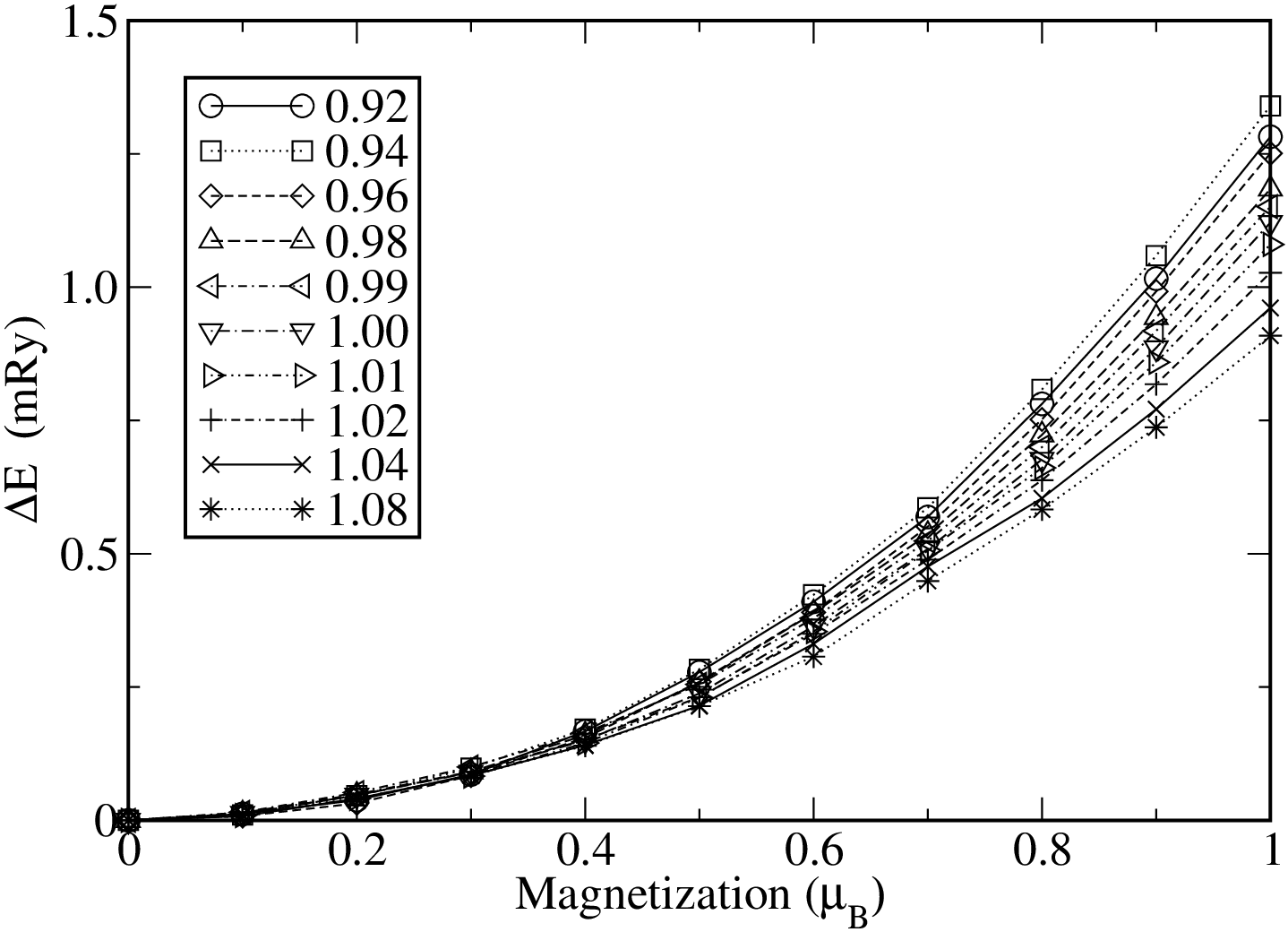}

\includegraphics[%
  clip,
  scale=0.35]{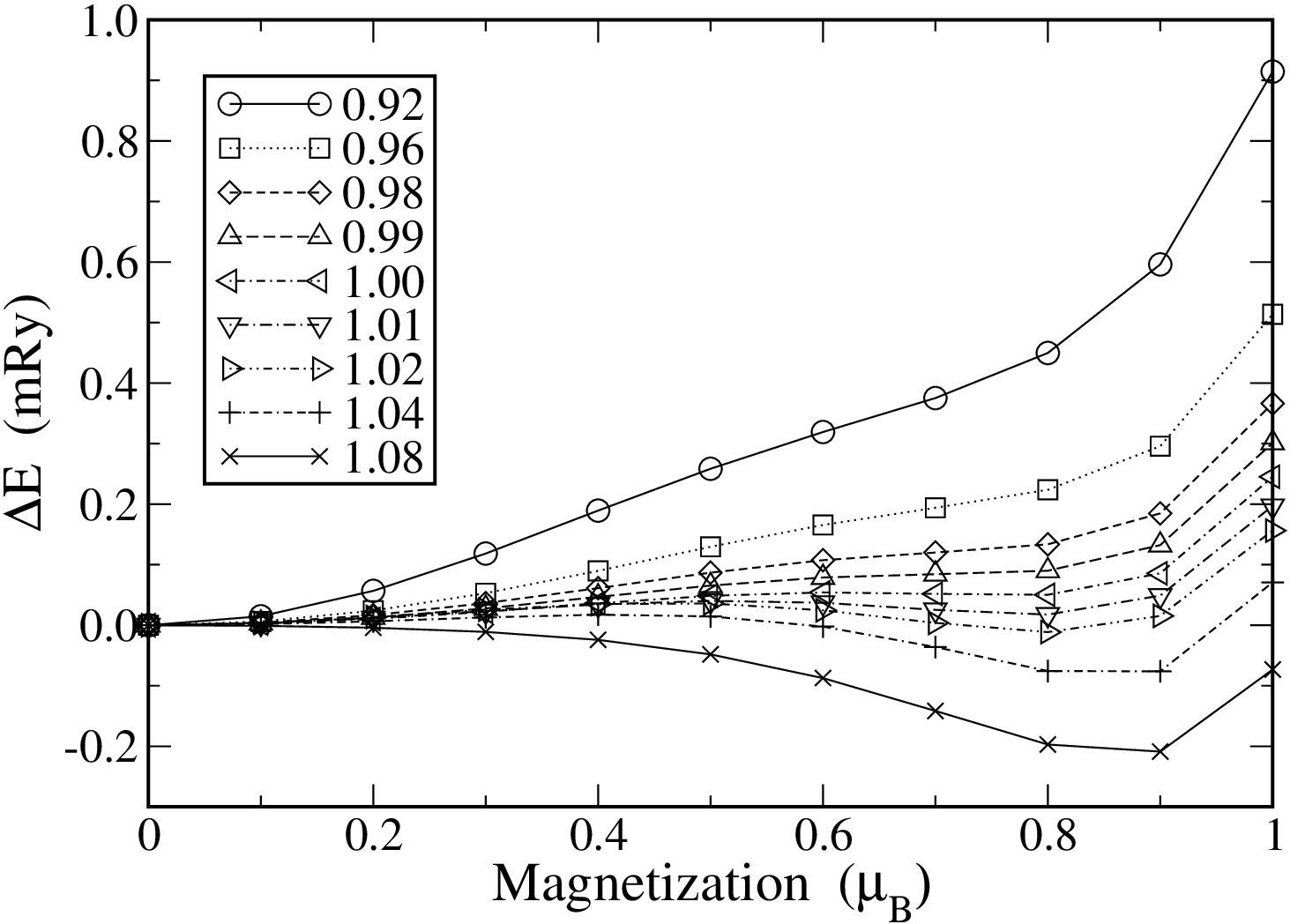}

\caption{\label{fsm-DE-mgcni3} The Magnetic energy $\Delta E(M)$ (in $mRy$)
as a function of magnetization $M$ (in $\mu_{B}$) for $MgCNi_{3}$
(upper panel) and $YBRh_{3}$ (lower panel), calculated as described
in the text.}
\end{figure}

As a cross check to the self-consistent calculations, as described
above, we have carried out fixed-spin moment calculations \cite{JPF-14-129}
to estimate the magnetic energy $\Delta E(M)=$$E(M)-E(0)$, where
$E(0)$ and $E(M)$ are the total energies for the paramagnetic phase
and ferromagnetic phase with a magnetic moment $M$ respectively,
for both $MgCNi_{3}$ and $YBRh_{3}$ at various lattice constants. 

In Fig.\ref{fsm-DE-mgcni3}, we show the calculated magnetic energy
$\Delta E(M)$ as a function of $M$ for various lattice constants
for both $MgCNi_{3}$ and $YBRh_{3}$. From Fig.\ref{fsm-DE-mgcni3},
it is clear that $MgCNi_{3}$ remains paramagnetic throughout, as
no minimum, other than for $M$=$0$, appears in $\Delta E(M)$ \emph{vs}.
$M$ curve. However, for $YBRh_{3}$ a second minimum appears in the
$\Delta E(M)$ \emph{vs}. $M$ curve, indicating a ferromagnetic transition.
Hence, both self-consistent and fixed-spin moment methods show a greater
propensity of magnetism in $YBRh_{3}$ than in $MgCNi_{3}$. The existence
of an enhanced magnetic behavior in $YBRh_{3}$is due to a weak hybridization
of the metal $d$ states with that of the non-metal $p$ states in
comparison with $MgCNi_{3}$. In $MgCNi_{3}$, one finds that the
$C$ $2p$ states and the $Ni$ $3d$ states hybridize strongly with
significant contribution to $N(E_{F})$ from the $C$ sub-lattice.

In conclusion, first-principles, density-functional-based calculations
find $YBRh_{3}$ to be closer to a magnetic instability than $MgCNi_{3}$
with respect to lattice constant. The enhanced propensity of magnetism
in $YBRh_{3}$ may be responsible for lowering the value of $T_{C}$,
in spite of high density of states at $E_{F}$. The superconductor
$YBRh_{3}$ thus falls in the same class as the refractory compounds
$VN$, where spin-fluctuations lower the $T_{C}$ drastically.

\end{document}